\documentclass{PoS}

\newcommand\srm{\scriptscriptstyle\rm}

\def\hbeta{H\,$\beta$}
\def\OIII{[O\,${\srm III}$]}
\def\OIIId{[O\,${\srm III}$]\,$\lambda\lambda$4959, 5007}
\def\MgII{Mg\,${\srm II}$}

\def\FeII{Fe\,${\srm II}$}

\def\civ{C\,${\srm IV}$}
\def\ciii{C\,${\srm III}$]}

\title{Microlensing to probe the quasar structure: spectrophotometry of Q2237+0305 and of J1131-1231{\thanks{Based on observations made with the ESO-VLT (Cerro Paranal, Chile; Proposals 073.B-0243(A\&B), 074.B-0270(A), 075.B-0350(A), 076.B-0197(A), 177.B-0615(A\&B), PI: F. Courbin; Proposal 71.A-0407(B\&E), PI: D.Sluse).}}}

\ShortTitle{Microlensing in Q2237+0305 and J1131-1231}

\author{\speaker{D. Sluse}$^a$, A. Eigenbrod$^a$, F. Courbin$^a$, D. Hutsem\'ekers$^b${\thanks {FNRS (Senior research associate)}}, J.-F. Claeskens$^b$, G. Meylan$^a$, E. Agol$^c$, J. Surdej$^b${\thanks {FNRS (honorary Research Director)}} \\
        $^a$ Laboratoire d'Astrophysique, Ecole Polytechnique F\'ed\'erale de Lausanne (EPFL) Observatoire de Sauverny, 1290 Versoix, Switzerland\\
        $^b$ Institut d'Astrophysique et de G\'eophysique, Universit\'e de Li\`ege, All\'ee du 6 Ao\^ut 17, B5C, B-4000 Sart Tilman, Belgium\\
        $^c$ Astronomy Department, University of Washington, Box 351580, Seattle, WA 98195, USA \\ 
        E-mail: \email{dominique.sluse@epfl.ch}}

\abstract{We present the main results of the first long-term spectrophotometric monitoring of the ``Einstein cross'' Q2237+0305 and of the single-epoch spectra of the lensed quasar J1131-1231. \\
From October 2004 to December 2006, we find that two prominent microlensing events affect images A \& B in  Q2237+0305 while images C \& D remain grossly unaffected by microlensing on a time scale of a few months. Microlensing in A \& B goes with chromatic variations of the quasar continuum. We observe stronger micro-amplification in the blue than in the red part of the spectrum, as expected for continuum emission arising from a standard accretion disk. Microlensing induced variations of the \ciii\ emission are observed both in the integrated line intensity and profile. Finally, we also find that images C \& D are about 0.1-0.3 mag redder than images A \& B. \\
The spectra of images A-B-C in J1131-1231 reveal that, in April 2003, microlensing was at work in images A and C. We find that microlensing de-amplifies the continuum emission and the Broad Line Region (BLR) in these images. Contrary to the case of Q2237+0305, we do not find evidence for chromatic microlensing of the continuum emission. On the other hand, we observe that the Balmer and \MgII\ broad line profiles are deformed by microlensing. These deformations imply an anti-correlation between the width of the emission line and the size of the corresponding emitting region. Finally, the differential microlensing of the \FeII\ emission suggests that the bulk of \FeII\ is emitted in the outer parts of the BLR while another fraction of \FeII\ is produced in a compact region. }

\FullConference{The Manchester Microlensing Conference: The 12th International Conference and ANGLES Microlensing Workshop\\
		 January 21-25 2008\\
		 Manchester, UK}

\begin{document}

\section{Introduction}



\begin{figure}[!ht]
\begin{centering}
\includegraphics[width=6.5cm]{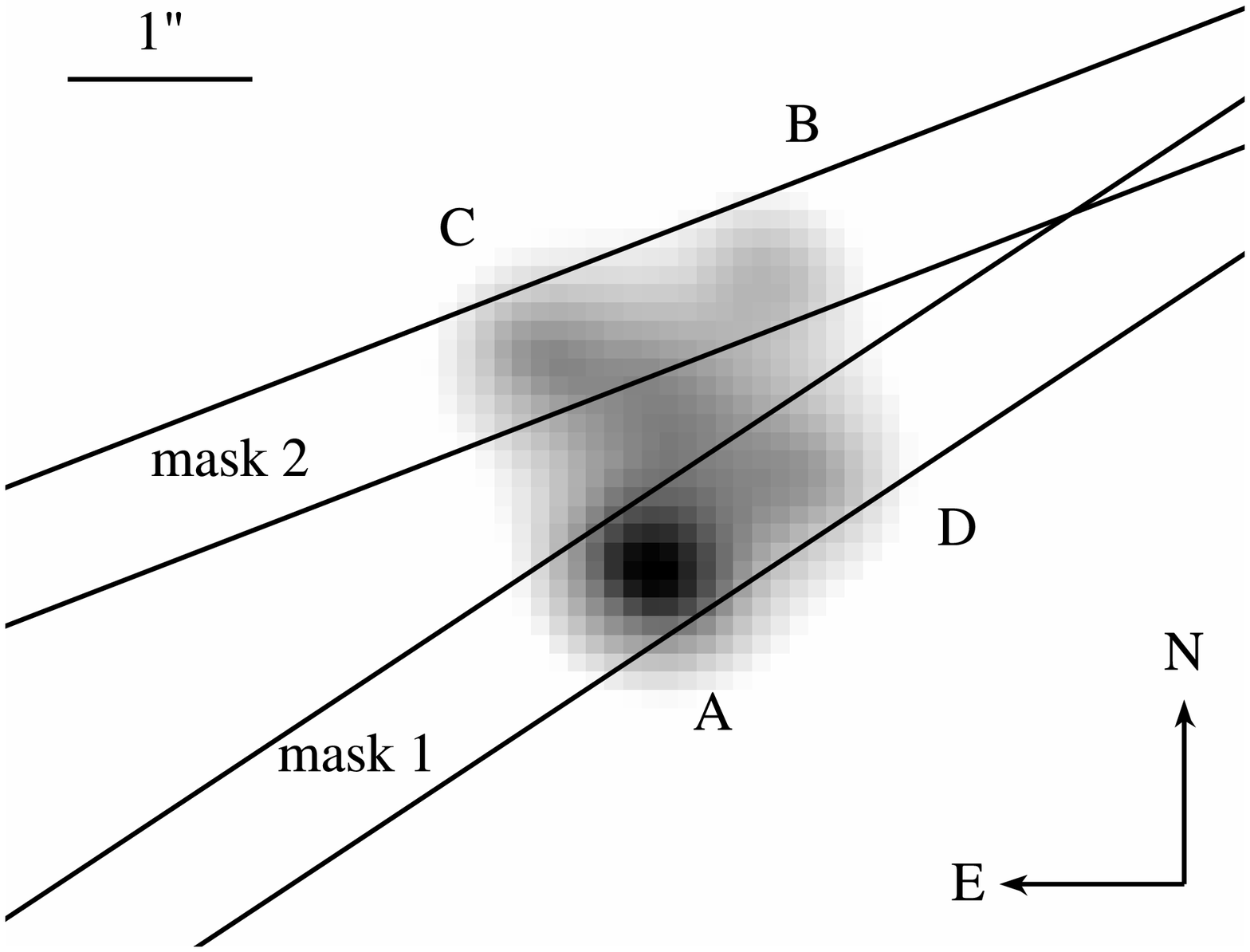}
\includegraphics[width=6.5cm]{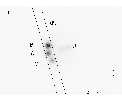}
\caption{{\it Left:} FORS1 R-band acquisition image of Q2237+0305 taken on 12 September 2005. Two different slits (0.7'' width) placed on images A \& D (mask 1) C \& D (mask 2) have been used. {\it Right:} FORS2 R-band image of J1131-1231 taken on 26 April 2003. The slit (1'' width) used is shown. The seeing on both images is of the order of 0.6''.}
\label{fig:aquisition}
\end{centering}
\end{figure}

So far, the effect of microlensing on the continuum and on the Broad
Line Region (BLR) of quasars has been investigated mainly
theoretically (e.g. Schneider \& Wambsganss 1990, Wambsganss \&
Paczynski 1991, Lewis et al. 1998, Abajas et al. 2002, Lewis \& Ibata
2004). In this contribution, we report the main results from a
spectrophotometric study of microlensing occurring in two
gravitationally lensed quasars: the ``Einstein cross'' Q2237+0305 and
J1131-1231 (Fig.~\ref{fig:aquisition}). The Einstein cross consists of
a $z_s= 1.695$ quasar gravitationally lensed into four images in a
cross-like pattern about the bulge of a nearby Sab galaxy ($z_l=
0.0394$). Due to the low redshift of the lensing galaxy, to the high
density of stars on the line of sight to the lensed images and to the
negligible time delay, this system is among the best ones to study
microlensing. We present hereafter some results of the first
long-term regular spectro-photometric monitoring campaign of this
object. The second target, J1131-1231, is one of the nearest
gravitationally lensed quasar. The lensing galaxy at $z_l = 0.295$
splits the light rays from the source at $z_s = 0.658$ into four
macro-images: three bright images (A-B-C) separated on the sky by
typically 1'' and a fainter component (D) located at 3.6'' from A
(Sluse et al. 2003). We study microlensing between the three brightest
images which also happen to have negligible differential time delays.


\section{Microlensing spectral signatures}

microlensing selectively (de-)magnifies regions with sizes
approximately smaller or equal to the projected size of the microlens
Einstein radius. Consequently, it induces chromatic variations of the
quasar continuum emission (Wambsganss \& Paczynski, 1991),
(de-)magnification of the whole BLR (Schneider \& Wambsganss 1990) or
deformations of the broad emission lines (e.g. Lewis
1998). Unfortunately, two other effects may also produce chromatic
variations in quasar spectra and could potentially be mistaken for
microlensing. First, each quasar image provides a snapshot of the
source at a different epoch due to the time delays between the lensed
images. If the source is intrinsically variable, this induces
significant spectral differences between lensed images. Our choice of
targets with differential time delays of the order of a day makes this
effect negligible. Second, differential reddening between lensed
images caused by the dust in the lensing galaxy might also exist. In
contrast with microlensing, this effect does not vary with time. We
show in Sect.~3 how the monitoring data of Q2237+0305
(Sect.~\ref{sec:Q2237}) allow us to disentangle between differential
extinction and microlensing in this system. For J1131-1231,
differential reddening is negligible (Sluse et al. 2006, 2007).



Three techniques allow to unveil and study microlensing spectral
signatures.  First, the ratio between individual spectra provides the
simplest diagnostic for the presence of chromatic difference between
the lensed images. This ratio is expected to be flat and equal to the
macro-magnification ratio if microlensing and/or differential
extinction are not at work. Another technique assumes macro and
micro-magnification factors and linearly combines the fluxes of 2
lensed images to unveil which fraction of the flux is
microlensed. Finally, a more quantitative approach is to decompose
the individual spectra into the multiple quasar emission components
(e.g. power law continuum, narrow and broad emission lines) and
compare the flux ratios in these individual components. We focus here
on the first approach. Sluse et al (2007) and Eigenbrod et al. (2008)
have developed the results of the last two techniques applied to
resp. J1131-1231 and Q2237+0305.

\section{Q2237+0305}
\label{sec:Q2237}

The Einstein cross has been spectro-photometrically monitored with the
FORS1 instrument (ESO Very Large Telescope) in the multi-object
spectroscopy (MOS) mode from October 2004 to November 2007. This
observational strategy allows us to get simultaneous observations of
the main target and of four stars used as reference point-spread
functions (PSFs). The 2D spectra of these stars are used to spatially
deconvolve the spectra of the target using the spectral version of the
MCS deconvolution algorithm (Magain et al. 1998, Courbin et
al. 2000). This allows to perform an accurate calibration of the
target spectra from one epoch to another (see Eigenbrod et al. 2008
for further details). We use the grism G300V that covers the
wavelength range $3900$\,\AA $< \lambda < 8200$\,\AA, corresponding to
the quasar rest-frame range 1400-3200\AA. Our spectra cover the \civ,
\ciii, \MgII\, broad emission lines, \FeII$_{\rm UV}$ and the quasar
UV continuum. We report here the results of the first 31-epoch
observations obtained from October 2004 to December 2006.

\begin{figure}[!ht]
\begin{centering}
\includegraphics[width=9cm]{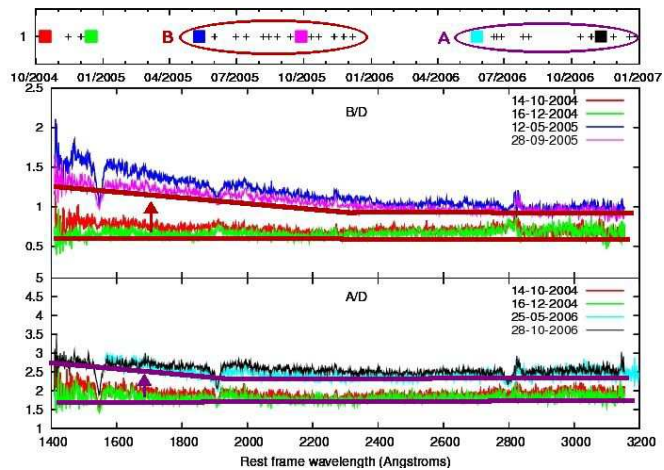}
\caption{Q2237+0305: Spectral ratios A/D (bottom panel) and B/D (middle panel) at 4 different epochs. In each panel, the green and red curves show the flux ratios observed in 2005. The 2 other curves show the spectral ratios calculated during the microlensing amplifications observed in 2005 (image B) and in 2006 (image A). The upper panel shows the time sampling of the spectro-photometric lightcurves. The colored squared correspond to the observational epochs displayed in the other panels.}
\label{fig:ratio}
\end{centering}
\end{figure}

\begin{figure}[!ht]
\begin{centering}
\includegraphics[width=9cm]{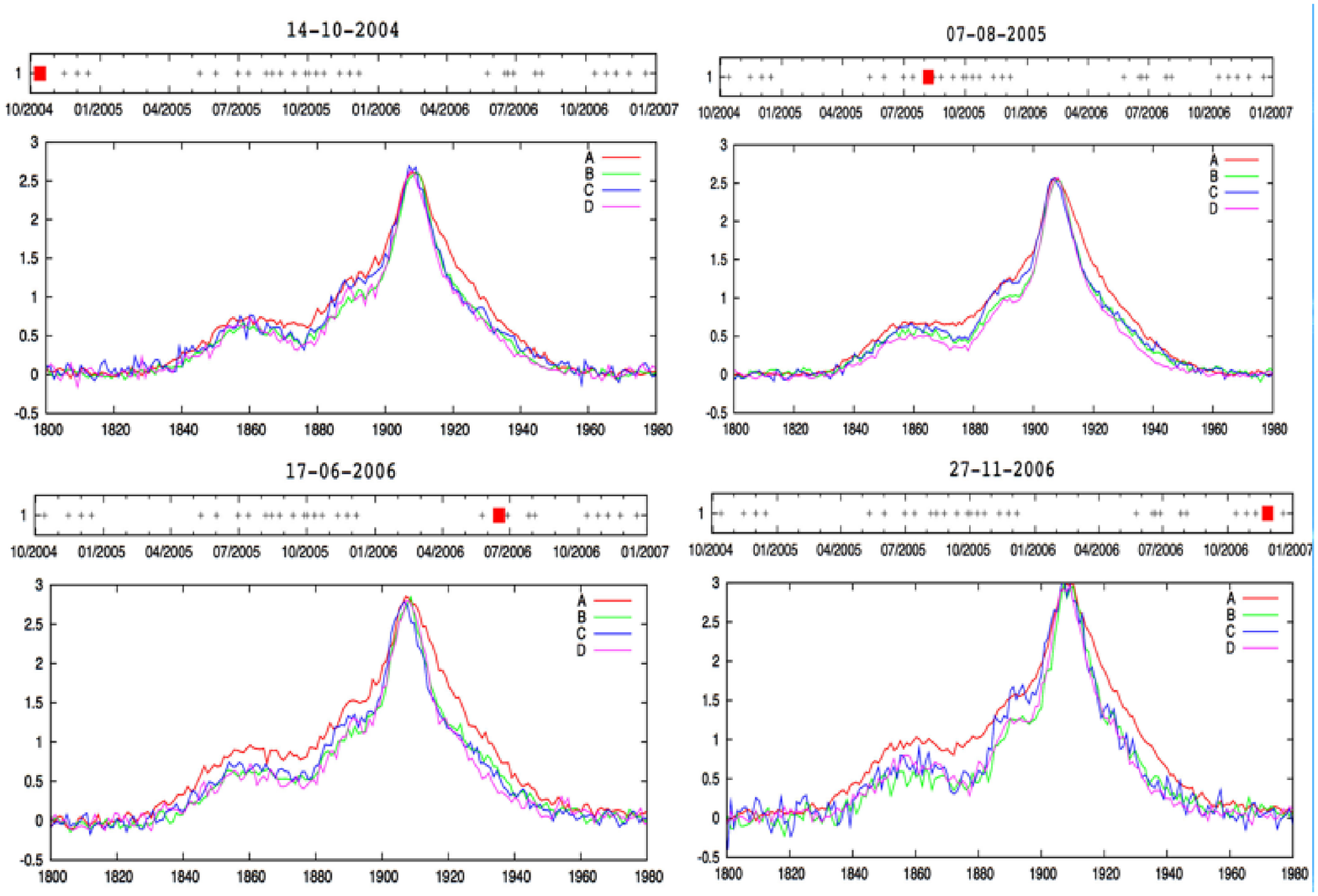}
\caption{Q2237+0305: Comparison of the \ciii\ emission line (normalized to the peak intensity of image A) observed in images A-B-C-D at 4 different epochs. }
\label{fig:ciii}
\end{centering}
\end{figure}

None of the spectral ratios (Fig.~\ref{fig:ratio}) between image pairs
is perfectly flat over the whole observing period, indicating that
microlensing occurs in all the lensed images. However, the flattest
and most stable spectral ratio is obtained for the pairs of images C
and D. This indicates that there is no differential reddening between
these two images and that no major wavelength dependent continuum flux
variations induced by microlensing occurred in images C or
D. However, the C/D ratio shows a clear imprint of the Broad Emission
Lines (BELs) indicating that the continuum of image C is significantly
more magnified than the BELs during the whole observing period. The
alternative scenario involving microlensing de-magnification of image
D is ruled out based on the comprehensive study of the flux ratios
between each possible image pair. Consequently, in the following, we
take image D as a reference image. The ratios A/D, B/D show a non zero
slope over the whole observing campaign, indicative of differential
reddening of image D with respect to images A and B. Using a Milky-way
extinction curve (Cardelli et al. 1989) with $R_v=3.1$, we find that
differential extinction in the range $0.1 {\rm mag} < A_V(D)-A_V(A)
\simeq A_V(D)-A_V(B) < 0.3$\ mag explains the observed slope. In
addition, significant changes are observed in the spectral ratio B/D
for the period May 2005-Dec 2005 (HJD 3500-3710) and in the ratio A/D
for the period May 2006-Dec 2006 (HJD 3880-4100). These variations are
the signature of two major chromatic microlensing events that
occurred in images B and A (Fig.~\ref{fig:ratio}).


Figure~\ref{fig:ratio} shows the A/D and B/D flux ratios at 4
different epochs, corrected for differential extinction. Microlensing
events observed in images A \& B are associated with a steepening of
the bluest part of the continuum that is more (micro-)amplified than
the redder part. This is a strong observational constraint that allows
us to determine the energy profile of the continuum emission, i.e. how
the size of the continuum emission region change with the wavelength
(Eigenbrod et al., in prep.). We also find a clear evidence that the
BLR is small enough to be significantly
microlensed. Figure~\ref{fig:ciii} shows the \ciii\, line profile at
four different epochs in the four lensed images, after normalization
to the peak intensity of image A. It is conspicuous that the line
profiles do not match perfectly: microlensing induces a global
brightening of the line profile and also slight deformations of the
line profile. The most striking difference observed in the line occurs
in image A where the broadest component of the \ciii\, emission line
is more microlensed than its core. This is consistent with a scheme
where the highest velocity component of the line is emitted in the
most compact region. On the other hand, the larger microlensed
induced variation of the \civ\, emission line indicates that \civ\, is
emitted in a more compact region than \ciii.


\section{J1131-1231}
\label{sec:J1131}

\begin{figure}[!ht]
\begin{centering}
\includegraphics[width=7.cm]{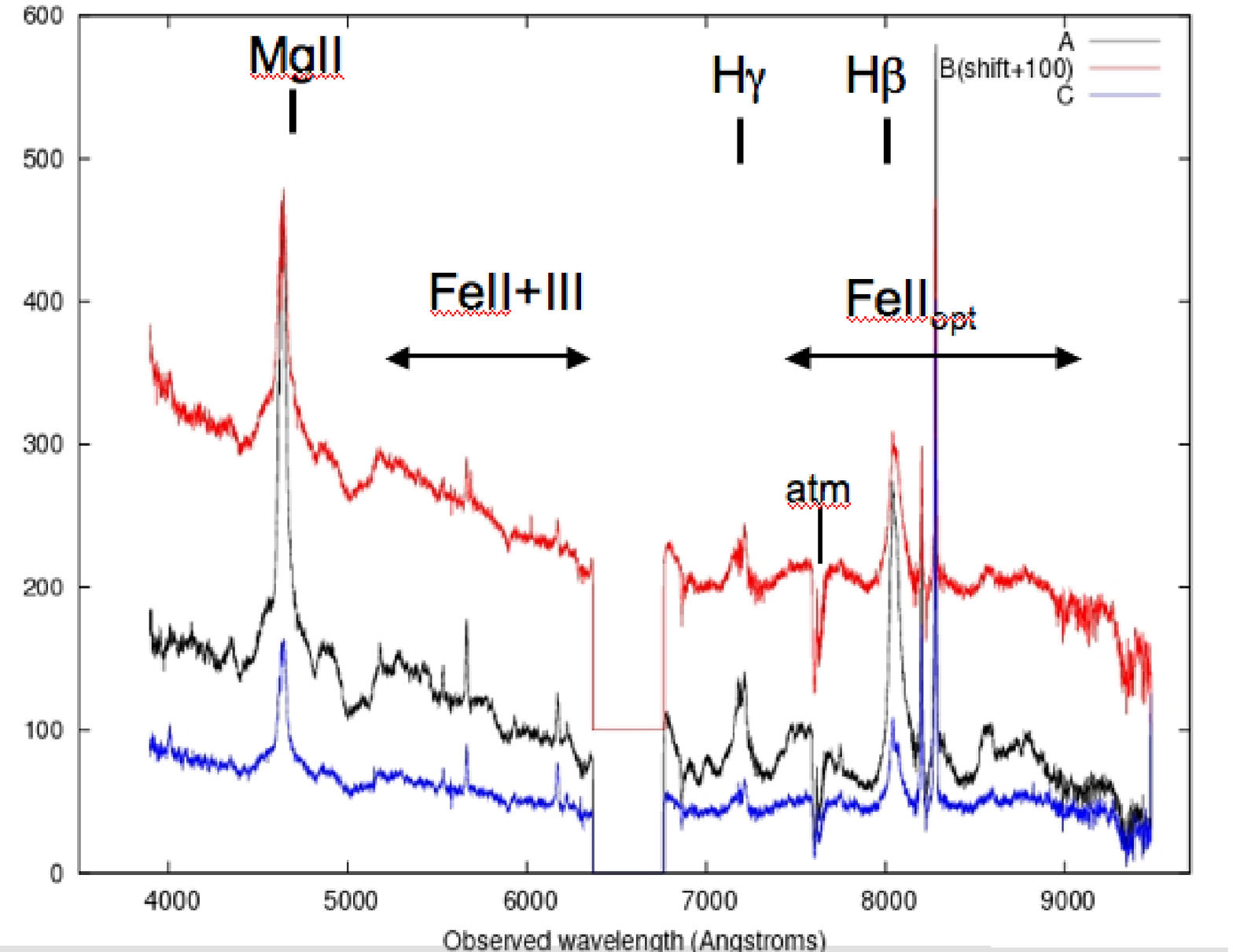}
\includegraphics[width=7.8cm]{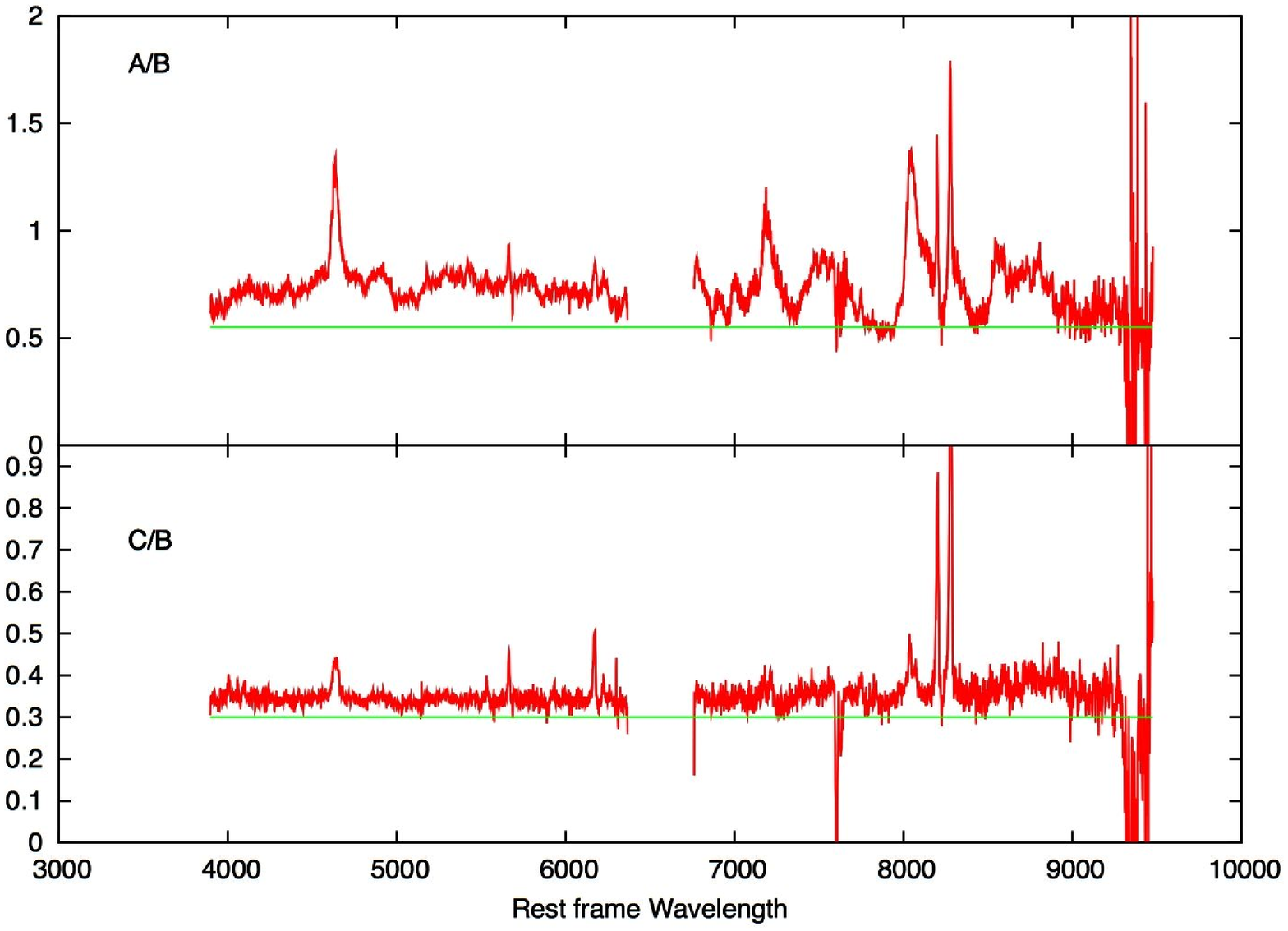}
\caption{{\it {Left:}} Extracted spectra of images B, A, C (from top to bottom) of J1131-1231. {\it {Right:}} Spectral ratios A/B (top) and C/B (bottom) in J1131-1231 after correction for the flux contamination by the quasar host galaxy as described in Sluse et al. (2007).}
\label{fig:ratioJ1131}
\end{centering}
\end{figure}

We present long-slit spectra of J1131-1231 obtained with the FORS2
instrument (ESO Very Large Telescope) on April 26th 2003. These data
consist of spectra obtained with the 1'' slit oriented along images
B-A-C. We used 2 grisms which cover the wavelength ranges $3890 $\,\AA\
$ < \lambda < 6280$\,\AA\, and $6760 $\,\AA\ $< \lambda < 8810$\,\AA\,
with a resolving power of 800 and 1500 at the respective central
wavelengths. This observational set up enables us to cover the (AGN)
rest-frame range 2500-5600~\AA.
Figure~\ref{fig:ratioJ1131} shows the extracted spectra
and indicates the main emission features.


As discussed in details in Sluse et al. (2007), our data suggest
microlensing de-magnification of images A and C.  This scenario is
further supported by optical variability and X-ray data presented in
Kochanek et al. (2006). Consequently, we consider image B as our
reference image, i.e. unaffected by
microlensing. Figure~\ref{fig:ratioJ1131} displays the spectral ratio
A/B and C/B.  The imprint of the QSO emission line spectra is
conspicuous in the A/B spectral ratio. This indicates that the flux
ratio is different in the continuum (e.g. 7900-7950~\AA), the BELs
(e.g. 8000-8170~\AA) and in the NELs (e.g. 8240-8320~\AA). Differences
in the pseudo-continuum (i.e. the AGN power law continuum+blended
\FeII~emission) is also apparent (e.g. 5100-6200~\AA). This is easily
understood if microlensing demagnifies more strongly the compact
region emitting the continuum than the larger line-emitting
regions. Although the continuum emission seems to be strongly
microlensed, we do not observe chromatic changes of the continuum
ratio over the studied wavelength range. Like the A/B continuum ratio,
the C/B continuum ratio is flat. However, the imprint of the broad
emission lines and of the narrow emission lines are well seen,
confirming different amounts of microlensing in the continuum, in the
broad and in the narrow emission lines. In addition, the C/B ratio
shows that the broad lines appear narrower than they are in the
individual spectra and that the pseudo continuum emission is nearly
invisible. Consequently, the flux ratio is similar in the broadest
component of the emission line, in the pseudo-continuum emission and
in the continuum emission in image C, indicating that these three
regions are affected by roughly identical amount of microlensing.

\begin{figure}[!ht]
\begin{centering}
\includegraphics[angle=-90, width=8cm]{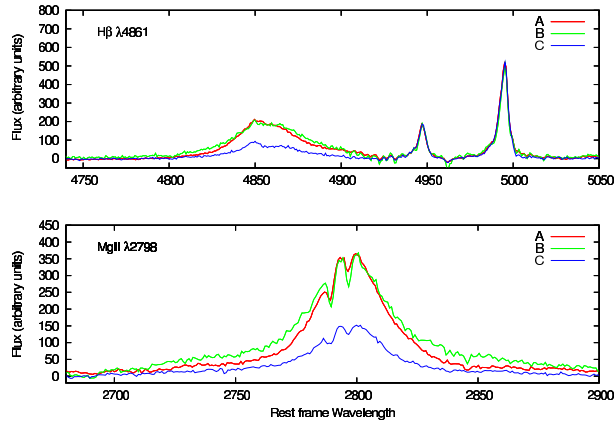}
\caption{Emission line spectra of images A-B-C of J1131-1231 normalized to the flux in the narrow continuum subtracted \OIIId\ emission. The upper panel shows the emission in the \hbeta\ region (4750-5050~\AA) and the bottom panel shows emission in the \MgII\, emission region (2660-2900~\AA). The continuum and pseudo-continuum emission have been subtracted following Sluse et al. (2007). }
\label{fig:emissionJ1131}
\end{centering}
\end{figure}


Because the region emitting the narrow \OIII\, emission is much larger
than the typical microlensing Einstein radius, we can reasonably
assume that the macro-amplification ratios are close to the flux
ratios measured in \OIII. Consequently, any mismatch between the
broad-emission lines after normalization to the \OIII\, flux is most
probably the signature of microlensing. We see in
Fig.~\ref{fig:emissionJ1131} that the BELs in image C are significantly
demagnified while only the broad component of the line is demagnified
in image A. This means that the wings of the broad emission lines are
emitted in a more compact region than the core of the line, as in the
Einstein Cross (Sect.~\ref{sec:Q2237}).

\begin{figure}[!ht]
\begin{centering}
\includegraphics[width=9cm]{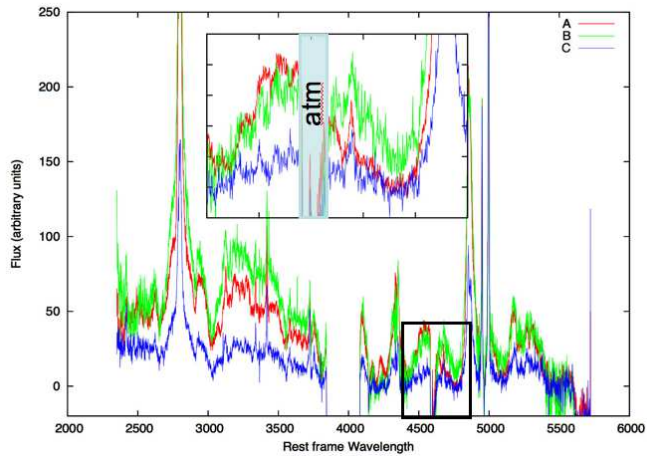}
\caption{Continuum subtracted spectra of images A-B-C of J1131-1231 normalized to the flux in the \OIIId\ emission. The inset graph is a zoom on the \FeII\ emitted in the range 4400-4900~\AA. }
\label{fig:FeIIJ1131}
\end{centering}
\end{figure}

We have also looked at microlensing of the \FeII\, emission in
J1131-1231. Figure~\ref{fig:FeIIJ1131} shows the spectra of images
A-B-C subtracted from the continuum emission and re-normalized to the
flux in the narrow \OIIId\, emission. For the \FeII\, emitted at
$\lambda < 3000$\,\AA\, and $\lambda > 5000$\,\AA, we see a behaviour
similar to the one observed in the \MgII\, and \hbeta\, emission: the
\FeII\, emission is only marginally microlensed in image A and
significantly microlensed in image C. On the other hand, the bulk of
the \FeII\, emission in the range $3000 $\,\AA\ $< \lambda <
5000$\,\AA\, show a different behaviour. Especially, the \FeII\,
emitted in the ranges 3080-3540\,\AA\ and 4630-4800\,\AA\ (panel inset
in Fig.~\ref{fig:FeIIJ1131}) seem significantly more de-amplified than
the BELs in image A, suggesting that it is emitted in more compact
regions.

\section{Conclusions} 

We have presented the main results of the first long term
spectro-photometric monitoring of the Einstein Cross Q2237+0305 and of
the single-epoch spectroscopic observations of the lensed quasar
J1131-1231. The microlensing spectral signatures observed in the
lensed images of these two systems confirm that quasar microlensing
is a powerful tool for studying the continuum and the broad line
emitting region of lensed quasars.

We find that all images of the Q2237+0305 are affected by some
microlensing during the observing period. Image D seems to be the
less affected. Our observations have monitored two important
microlensing brightening episodes in image B and A on resp. May
2005-Dec 2005 (HJD 3500-3710) and May 2006-Dec 2006 (HJD
3880-4100). For each of these events, the continuum becomes bluer when
the image gets brighter as expected from microlensing magnification
of an accretion disk. In addition to the microlensing induced
chromatic variations of the quasar continuum we have also observed
time-independent reddening of images C and D with respect to A and
B. We estimate that the amount of differential extinction between
pairs of quasar images is in the range 0.1-0.3 mag, with images C and
D being the most reddened. We also report microlensing induced
variations of the BELs affecting both the integrated line intensities
and profiles. The microlensing of the BLR is the strongest in image A
where we observe that the broad component of the broad lines
is more microlensed than the core. This is compatible with an
anti-correlation between the line width and the size of the
corresponding emitting region. Finally, we observe that the continuum
of images C and A are significantly more magnified than the BELs
during the whole observing period, indicating that long-term
microlensing is at work in those two images.

The analysis of the single-epoch spectra of images A-B-C of J1131-1231
has revealed microlensing de-amplification of the quasar images A
and C. In these two cases, the continuum and the BELs are
microlensed. However, a larger fraction of the BELs is microlensed
in image C, indicating that the Einstein radius of the microlens is
larger for that image. Microlensing of the BELs confirms that the
size of the emission line region is anti-correlated with the FWHM of
the corresponding line component. Contrary to the case of Q2237+0305,
we do not find evidence for chromatic microlensing of the
continuum. The most interesting result concerns the microlensing of
the \FeII\ emission. On one hand, we find that a large fraction of the
near UV and optical \FeII\, emission takes place in a region similar
to the outer parts of the Balmer Line Emitting Region. On the other
hand we find that the \FeII\ emitted in the rest-frame ranges
4630-4800 \& 3080-3540\AA\ likely arises from a more compact region
possibly similar in size to the region emitting the very broad \MgII\
emission.

Our observations have demonstrated that the BLR of the quasars
Q2237+0305 and J1131-1231 is small enough to be significantly
microlensed. Because of the different source redshifts, these two
targets allow to study very different emitting regions in the optical
range. Microlensing in Q2237+0305 allows to study the near UV
continuum emission, the \FeII$_{\rm UV}$, and the high ionisation
\ciii, \civ\ broad emission lines while microlensing in J1131-1231
allows to probe the rest-frame continuum emission in the range
2500-6000 \AA, the \FeII$_{\rm opt}$, the low ionisation \MgII, and
the Balmer broad emission lines. The spectro-photometric monitoring of
Q2237+0305 demonstrates the great asset of the time information in the
study of quasar microlensing. We have began a 2 year
spectrophotometric monitoring of J1131-1231 with the VLT to track the
microlensing induced deformation of the BELs in this system. The
spectro-photometric monitoring data gathered for Q2237+0305 and soon
for J1131-1231 will allow to put sharp upper limits on the size of the
continuum emission region and on the geometry of the BLR (for
different ionization levels) using state-of-the-art inverse
ray-shooting simulations.

{\large{\bf {Acknowledgements:}}} This work is supported by the Swiss
National Science Foundation, by ESA PRODEX under contract PEA
C90194HST and by the Belgian Federal Science Policy Office.

\end{document}